\documentstyle[psfig]{l-aa}
\begin{document}
\topmargin 50pt

\thesaurus{9(2.14.1, 9.01.1, 9.13.2, 10.01.1)}

\title{Interstellar $^{36}$S: a probe of s-process nucleosynthesis
      \thanks{Based on observations with the Swedish-ESO Submillimetre
              Telescope, SEST, and the 12-m radiotelescope of the National
              Radio Astronomical Observatory, NRAO.
              NRAO is a facility of the National Science Foundation operated
              under cooperative agreement by Associated Universities Inc.} }

\author{R.~Mauersberger\inst{1} \and C.~Henkel\inst{2} \and
        N.~Langer\inst{3} \and Y.-N.~Chin\inst{4,5}}

\offprints{R. Mauersberger}

\institute{
   Steward Observatory, The University of Arizona, Tucson, AZ 85721, U.S.A.
\and
   Max-Planck-Institut f\"ur Radioastronomie,
   Auf dem H\"ugel 69, D-53121 Bonn, Germany
\and
   Max-Planck-Institut f\"ur Astrophysik,
   Karl-Schwarzschild-Str. 1, D-85740 Garching, Germany
\and
   Institute of Astronomy and Astrophysics, Academia Sinica,
   P.O., Box 1-87, Nankang, 115 Taipei, Taiwan
\and
   Radioastronomisches Institut der Universit\"at Bonn,
   Auf dem H\"ugel 71, D-53121 Bonn, Germany
}
\date{received ; accepted}
\maketitle

\begin{abstract}
   The first detection of a $^{36}$S-bearing molecule
   in interstellar space is reported.
   The $J$=$2-1$ and $3-2$ transitions of C$^{36}$S have been observed
   toward eight Galactic molecular hot cores.
   From a comparison with other optically thin isotopic species of CS,
   the abundance ratio of $^{34}$S/$^{36}$S is 115$(\pm 17)$.
   This is smaller than the solar system ratio of 200 and
   supports the idea that $^{36}$S is, unlike the other stable
   sulfur isotopes, a purely secondary nucleus that is produced by
   s-process nucleosynthesis in massive stars.


\end{abstract}

\section{Introduction}

   Sulfur possesses four stable isotopes, $^{32}$S,
   $^{33}$S, $^{34}$S, and $^{36}$S.
   In the solar system, the abundance ratios are
   95.02:0.75:4.21:0.021 (Anders \& Grevesse 1989).
   For the interstellar medium, Chin et al.\ (1996) determine abundance ratios
   of 24.4$(\pm 5.0)$ for $^{32}$S/$^{34}$S,
   and 6.3$(\pm 1)$ for $^{34}$S/$^{33}$S.
   While no variation in the $^{34}$S/$^{33}$S isotope ratio is found,
   the $^{32}$S/$^{34}$S ratio may increase with galactocentric radius.
   In the local interstellar medium (ISM) $^{32}$S,
   $^{33}$S and $^{34}$S abundance ratios are
   not drastically different from those in the solar system.
   For $^{36}$S, no interstellar data have been obtained so far.

   $^{32}$S, $^{33}$S, and $^{34}$S are mainly primary products of
   oxygen burning (Woosley et al.\ 1973, Chin et al.\ 1996);
   however, especially the $^{34}$S yield retains a substantial sensitivity
   to the metallicity (Woosley \& Weaver 1995).
   $^{36}$S is thought to be a purely secondary isotope produced by
   neutron captures on the primary sulfur seeds during helium and
   carbon burning (Thielemann \& Arnett 1985, Langer et al.\ 1989).
   This isotope is, hence, formed in a process significally
   different from that of the other sulfur or CNO isotopes accessible by
   means of molecular spectroscopy at cm- and mm-wavelengths.

   We have observed eight Galactic hot cores in two rotational lines
   of C$^{36}$S in order to determine the abundance of $^{36}$S relative
   to other sulfur isotopes and to relate this isotopic abundance pattern
   to that of the local Galaxy 4.6\,Gyrs ago, i.e.\ the solar system value.
\begin{figure*}
   \vspace*{-24mm}
   \psfig{figure=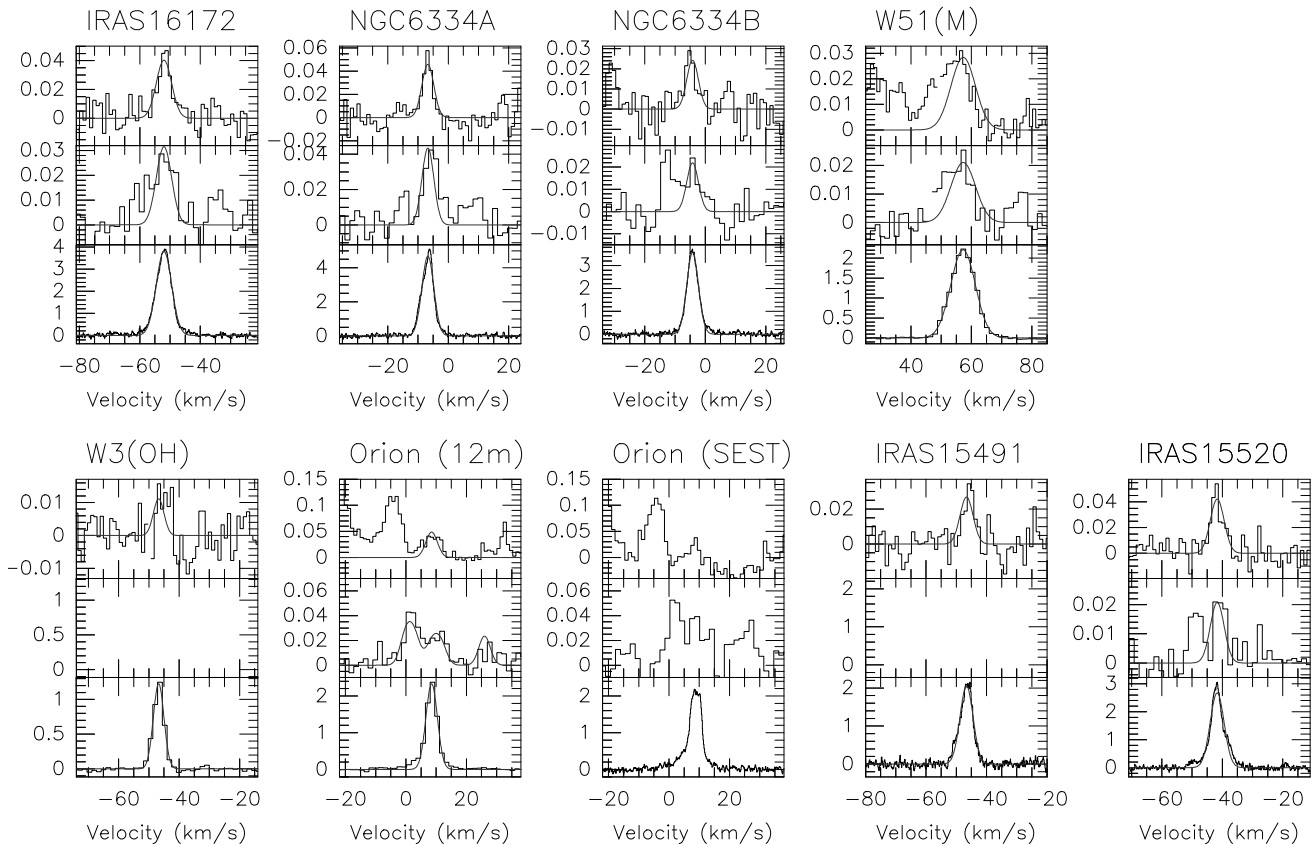,width=18.0cm,angle=270}
   \vspace*{10mm}
   \caption[]
           {The observed profiles of the lines of C$^{34}$S $2-1$
            (lower frames), C$^{36}$S $2-1$ (mid frames), and C$^{36}$S $3-2$
            (upper frames).
            The velocity is with respect to the Local Standard of Rest (LSR),
            the intensity scale is $T_{\rm MB} (\rm K)$.
            Also shown are Gaussian fits to the lines.
            For Orion, a simultaneous fit of the C$^{36}$S $2-1$ line
            with a series of CH$_3$CCCN lines is shown.}
 \label{fig:C36S-profiles}
\end{figure*}

\section{Observations and results}

   The $J=2-1$ lines of $^{13}$C$^{32}$S, C$^{34}$S, and C$^{36}$S
   (95.016711\,GHz, Lovas 1984), and the $J=3-2$ transitions of C$^{34}$S and
   C$^{36}$S (142.522797\,GHz, Lovas 1984;
   for the other line frequencies see Lovas 1992) were observed
   in January and June 1996 with the 12\,m radiotelescope of the
   National Radioastronomy Observatory (NRAO) on Kitt Peak toward
   W3(OH), Orion-KL and W51(M).
   The same transitions as well as $^{13}$C$^{32}$S $3-2$ were observed
   in March and June 1996 toward IRAS\,15491$-5426$, IRAS\,15520$-$5234,
   IRAS\,16172$-5028$, NGC\,6334A, NGC\,6334B and also Orion-KL using the
   15\,m Swedish ESO Submillimetre Telescope SEST on La Silla.
   In all cases, weather conditions were good or excellent.
   SIS receivers were used with sideband rejections of 20\,dB or better.
   A main-beam brightness temperature ($T_{\rm MB}$) scale
   was established by a chopper wheel method.

\paragraph{12\,m NRAO observations.}
   The FWHP beamwidth of the 12\,m antenna was 65$''$ for the $2-1$ transitions,
   and 41$''$ for the $3-2$ transitions.
   From maps of continuum sources, we estimate the pointing to
   be correct within 10$''$.
   The focus was determined after sunset and after sunrise.
   As backends, we employed a filterbank with 2$\times$128 contiguous channels.
   Each half of the filterbank was connected to one of two orthogonal
   polarization channels of the receiver used.
   The channel spacing of 500\,kHz corresponds to a velocity resolution
   of $\sim 1.6\,\rm km\,s^{-1}$ for the $J=2-1$ transitions
   and 1.1 km\,s$^{-1}$ for the $J=3-2$ transitions.
   All spectral lines were measured in position switch mode,
   the reference position being at $+30'$ in Right Ascension relative
   to the source position; 30s of integration time on the reference
   position were followed by 30\,s on the source.

\paragraph{15\,m SEST observations.}
   The $J=2-1$ and $3-2$ lines of each isotopic species were observed
   simultaneously with two single channel SIS receivers.
   At $\lambda=3\,\rm mm$, the beam has a width of 51$''$, and at 2\,mm,
   the width is 36$''$.
   The pointing was estimated to be accurate within 5$''$.
   A dual beam switch mode with reference positions offset
   by $\pm 11'40''$ in azimuth was employed.
   The backend consisted of two acousto optical spectrometers
   with 46\,kHz channel separation.
   For the presentation of the C$^{36}$S data,
   the resolution was reduced to 0.5\,MHz.

\paragraph{Results.}
   We subtracted linear baselines from all profiles.
   The resulting spectra of C$^{34}$S $J=2-1$ and C$^{36}$S
   $2-1$ and $3-2$ are shown in Fig.\,\ref{fig:C36S-profiles}.
   The profiles for the C$^{34}$S $3-2$ and the $^{13}$C$^{32}$S $2-1$ and
   $3-2$ transitions have, within the noise, the same lineshape
   as C$^{34}$S $2-1$, and are therefore not displayed.
   The spectra toward Orion show that the calibration scale of the
   12\,m telescope and the SEST are, within the noise, the same.
   Confidence in our results is strengthened by the fact that both
   telescopes show the same spectral features.
   In the following we will use the 12\,m data for Orion,
   which have a better signal-to-noise ratio.
   Spectral features at the frequency of the $J=2-1$ and $3-2$ lines
   of C$^{36}$S have been detected toward all sources observed (toward W3(OH)
   and IRAS\,15491 we only searched for the C$^{36}$S $3-2$ line).
   We have fitted Gaussians to all the lines.
   The results are displayed in Table\,\ref{tbl:C36S-gauss}.

\begin{table*}
   \caption[]{Line Parameters from Gaussian fits}
 \label{tbl:C36S-gauss}
   \begin{tabular}{l r r r r r r r r}
   \hline
   Source$^a$   & $D_{\rm GC}$ & $J-J'$ & $I(\rm C^{36}S)$ & $v_{\rm LSR}^b$
                & $\Delta v_{1/2}^b$
                & $\frac{I({\rm C^{34}S})}{I(\rm {C^{36}S})}$
                & $\frac{I(^{13}{\rm C^{32}S})}{I(\rm {C^{36}S})}$
                & $\frac{^{12}{\rm C^{32}S}}{\rm {C^{36}S}}^c$ \\
                & (kpc)        &        & mK\,km\,s$^{-1}$ & km\,s$^{-1}$
                & km\,s$^{-1}$
                & & & \\
   \hline
   W3(OH)$^d$   &  10.4  &  $3-2$  &   49(13)  & $-$46.8 & 4.2
                &  181(49)  &         &            \\
   Orion-KL$^d$ &   9.0  &  $2-1$  & $\sim 83$ &     8.5 & 4.5
                &  114(50?) & 46(20?) & 3450(1500) \\
                &        &  $3-2$  & 233(16) &         &
                &  104(7)   &         &            \\
   IRAS\,15491  &   6.1  &  $3-2$  &  146(20)  & $-$46.6 & 5.1
                &  108(15)  & 57(7)   & 3020(370)  \\
   IRAS\,15520  &   6.2  &  $2-1$  &  120(30)  & $-$41.7 & 5.4
                &  128(32)  & 68(17)  & 3680(920)  \\
                &        &  $3-2$  &  243(20)  &         &
                &  133(11)  & 77(6)   & 4170(320)  \\
   IRAS\,16172  &   5.6  &  $2-1$  &  200(30)  & $-$52.0 & 6.0
                &  121(18)  & 60(9)   & 2980(450)  \\
                &        &  $3-2$  &  257(30)  &         &
                &  119(14)  & 64(7)   & 3170(350)  \\
   NGC\,6334A   &   7.0  &  $2-1$  &  216(30)  &  $-$6.7 & 4.7
                &  108(15)  & 52(7)   & 3120(420)  \\
                &        &  $3-2$  &  228(30)  &         &
                &  154(20)  & 83(11)  & 4990(660)  \\
   NGC\,6334B   &   7.0  &  $2-1$  &  106(30)  &  $-$4.2 & 4.5
                &  163(46)  & 76(22)  & 4570(1320) \\
                &        &  $3-2$  &  117(20)  &         &
                &  156(26)  & 75(13)  & 4510(780)  \\
   W\,51(M)$^d$ &   6.5  &  $2-1$  & 214(23) &    57.2 & 9.6
                &  105(11)  & 43(5)  & 2420(280)   \\
                &        &  $3-2$  & 290(40) &         &
                &  104(15)  &        &             \\
   \hline
   \end{tabular}\\
\footnotesize{
   a) Coordinates used (B1950.0):
      W3(OH): $2^{\rm h}23^{\rm m}16\fm 8$, $\delta_{1950}=61^{\rm o}38'57''$,
      Orion-KL: $5^{\rm h}32^{\rm m}46\fm 7$, $-5^{\rm o}24'24''$,
      IRAS\,15491: $15^{\rm h}49^{\rm m}13\fs 0$, $-54^{\rm o}26'30''$,
      IRAS\,15520: $15^{\rm h}52^{\rm m}00\fs 1$, $-52^{\rm o}34'26''$,
      IRAS 16172: $16^{\rm h}17^{\rm m}13\fs 3$, $-50^{\rm o}28'14''$,
      NGC\,6334A: $17^{\rm h}17^{\rm m}32\fs 0$, $-35^{\rm o}44'05''$,
      NGC\,6334B: $17^{\rm h}17^{\rm m}34\fs 0$, $-35^{\rm o}42'08''$,
      and W51(M): $19^{\rm h}21^{\rm m}26\fm 4$, $14^{\rm o}24'42''$;
   b) v$_{\rm LSR}$ and $\Delta v_{\rm 1/2}$ have been fixed in the fit
      using the values determined from C$^{34}$S $2-1$.
   c) see text;
   d) measured with the 12\,m telescope
}
\end{table*}

\section{The identification of interstellar C$^{36}$S}

   From Fig.\,\ref{fig:C36S-profiles} it is evident that close to
   the $2-1$ line frequency of C$^{36}$S there is molecular emission
   from another line toward some of the sources, such as Orion,
   IRAS\,15520, NGC\,6334B and possibly toward W51(M).
   The $3-2$ line appears to be less contaminated except
   toward Orion and W51(M).
   It is known that chemically younger molecular clouds such as
   Orion tend to have a more complex chemistry than older clouds where
   mainly simple molecules are observed (Helmich et al.\ 1994).
   This can explain the diffences in line blending.

   Close to the C$^{36}$S $2-1$ transition there are several $K$ lines of the
   $J=23-22$ transition of CH$_3$C$_3$N (methyl cyanoacetylene; Lovas 1984).
   This symmetric top molecule has been previously detected
   toward TMC-1 (Broten et al.\ 1984).
   Assuming that the population of the corresponding levels can be described by
   a Boltzmann law with a kinetic temperature of 90\,K
   (derived from the very similar CH$_3$CN, Cummins et al.\ 1983)
   and using the line strengths and level energies from Lovas (1984),
   we made a Gaussian fit fixing the estimated relative intensities of the
   CH$_3$C$_3$N lines for Orion, where line contamination is worst.
   The resulting estimate for the intensity of the C$^{36}$S $2-1$
   line is given in Table\,\ref{tbl:C36S-gauss}.

   The only line in the Lovas (1984) line catalog potentially contaminating
   the $3-2$ line of C$^{36}$S is a transition of CH$_2$CHCN (vinyl cyanide;
   $\nu=$ 142.527\,GHz) displaced by 4.5\,MHz
   (9.4\,km\,s$^{-1}$) from C$^{36}$S.
   Vinyl cyanide is known to be abundant in Orion but probably not in
   massive star forming regions with an ``older'' chemistry.
   The frequency separation is sufficiently high for most of the
   soures investigated to separate it clearly from the C$^{36}$S emission.

   We have further evidence that our identification of C$^{36}$S
   is indeed correct.
   As it is expected for the optical thin case, the $(3-2)/(2-1)$ line ratios
   are similar for each of the optically thin species, including C$^{36}$S.
   Also there is no large variation of the line ratios between different
   isotopomers of a given transition for different sources.
   We conclude that our identification of C$^{36}$S is justified.
   The line parameters are, however, best determined in sources with an
   ``old'' chemical composition (i.e.\ less complex molecules),
   such as IRAS\,16172, NGC\,6334A, and IRAS\,15520.

\section{The isotopic abundances of sulfur}

   We concentrate here on the $^{34}$S/$^{36}$S abundance ratio
   since $^{12}$C$^{32}$S lines tend to be moderately optically thick in
   Galactic star-forming regions
   (e.g.\ Frerking et al.\ 1980, Linke \& Goldsmith 1980, Chin et al.\ 1996);
   $^{13}$C$^{32}$S would involve a double isotope ratio and
   C$^{33}$S is less easy to interpret due to its hyperfine structure.
   There is only a slight Galactic gradient
   in the $^{32}$S/$^{34}$S abundance ratio from the data of
   Chin (1995) and Chin et al.\ (1996).

\begin{figure}
   \vspace*{-8mm}
   \psfig{figure=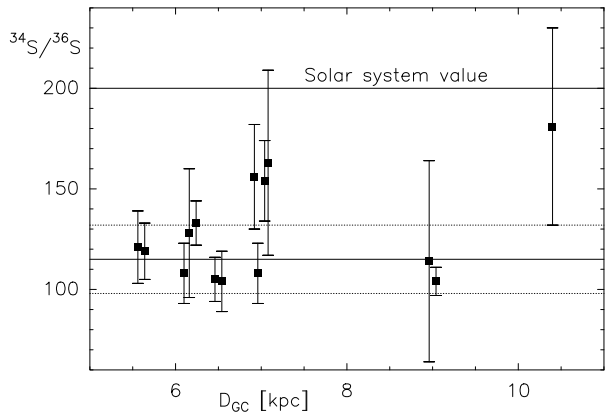,width=8.5cm,angle=270}
   \caption[]
           {The derived $^{34}$S/$^{36}$S abundance ratios from the $2-1$ and
            $3-2$ lines as a function of the galactocentric distance.
            The thick and the dashed lines show the weighted mean of the
            measurements and the standard deviation of the data.
            Also shown is the solar system value of 200.}
 \label{fig:C36S-abundances}
\end{figure}

   Since C$^{34}$S and C$^{36}$S are most probably both optically thin and
   since both have very similar molecular constants and were measured with
   the same telescopes we can assume that the ratio of integrated intensities
   (see Table \ref{tbl:C36S-gauss}) for a given rotational transition
   equals the respective abundance ratio.
   From the data in Table\,\ref{tbl:C36S-gauss},
   we obtain a weighted mean abundance ratio of $^{34}$S/$^{36}$S=115($\pm 17$).
   The standard deviation given is the square root of the weighted
   average variance of the data (Bevington \& Robinson 1992).
   Since we cannot exclude, at the present time, that part of the scatter
   in the data is caused by a Galactic gradient or by less systematic variations
   from source to source, we believe that this reflects the
   uncertainties better than the standard deviation of the weighted mean.
   The unweighted mean is 128($\pm 25$).
   The interstellar value of $^{34}$S/$^{36}$S is considerably smaller than
   the solar system ratio of 200 (see Fig.\,\ref{fig:C36S-abundances}).
   There is no systematic difference between the ratios obtained
   from the $2-1$ and $3-2$ lines.
   Due to the scatter of our data and the narrow range in galactocentric
   distances it is premature to discuss a Galactic abundance gradient.
   Using
\begin{equation}
   \frac{{\rm ^{32}S}}{\rm {^{36}S}}=\frac{I(^{13}{\rm C^{32}S})}
        {I(\rm {C^{36}S})} \frac{^{12}\rm C}{^{13}\rm C} \nonumber
\end{equation}
   and the $^{12}$C/$^{13}$C relation with galactocentric radius from
   Wilson \& Rood (1994), we obtain as the weighted mean value from
   our data $^{32}$S/$^{36}$S=3280$(\pm 760)$.
   Also for these isotopes, the interstellar ratio is smaller
   than the solar system value (4520).

   According to the metallicity dependent yields of $^{34}$S in massive stars
   (Woosley \& Weaver 1995), this isotope, which is produced during oxygen
   burning, has a character intermediate to primary and secondary isotopes.
   $^{36}$S, on the other side, is produced as a purely secondary isotope
   during s-process nucleosynthesis in massive stars, predominantly in helium
   and carbon burning (Thielemann \& Arnett 1985, Langer et al.\ 1989),
   with a possible (also secondary) contribution from AGB stars.
   Only a small fraction of the produced $^{36}$S is destroyed in later burning
   phases and during the supernova explosion (S.E.\ Woosley, priv.\ comm.).
   According to the comprehensive massive star models of
   Woosley \& Weaver (1995), the ratio of the production factors
   (i.e.\ output vs.\ input) of $^{34}$S over $^{36}$S in
   $15 - 25\,\rm M_{\odot}$ stellar models drops roughly from 10 to 1
   when the stellar initial metallicity is increased from
   $\rm Z_{\odot}/10$ to $\rm Z_{\odot}$.

   A larger solar $^{34}$S/$^{36}$S ratio than the corresponding present
   ISM value is thus in agreement with stellar $^{36}$S yields increasing with
   time and simultaneously constant $^{34}$S yields during Galactic evolution,
   i.e.\ a larger ISM value 4.6\,Gyr ago compared to today.
   In fact, a similar trend is found in two other isotope ratios which are
   clearly primary vs.\ secondary, namely for $^{12}$C/$^{13}$C
   ($\odot$: 89, local ISM: 70) and $^{16}$O/$^{17}$O ($\odot$: 2700,
   local ISM: 1900; cf.\ Henkel et al.\ 1995).
   This analogy suggests a positive Galactic $^{34}$S/$^{36}$S gradient,
   similar to those observed for the mentioned carbon and oxygen isotope ratios.
   This can be confirmed or rejected by observations of C$^{36}$S and
   C$^{34}$S in sources covering a larger range of galactocentric radii.

\section{Conclusions}

   For the first time we have detected a molecule containing $^{36}$S in space.
   C$^{36}$S could be measured in two rotational transitions
   and toward eight molecular hot cores.
   The interstellar $^{34}$S/$^{36}$S abundance ratio of
   115 $(\pm 17)$ is lower than in the solar system (200).

   This is in agreement with the idea that $^{36}$S is synthesized
   as a secondary isotope by s-process nucleosynthesis in massive stars.

\begin{acknowledgements}
   R.M.\ and N.L.\ were supported by Heisenberg fellowships by Deutsche
   Forschungsgemeinschaft (DFG grants Ma~1450/2-1 and La~587/8-2).
   It is a pleasure to thank S.E.~Woosley for illuminating discussions.
\end{acknowledgements}


\begin{thebibliography}{}
 \bibitem[]{}
   Anders, E., Grevesse, N., 1989, Geochim. et Cosmochim. Acta, 53, 197
 \bibitem[]{}
   Bevington, P.R., Robinson, D.K., 1992, Data reduction and error analysis
     for the physical sciences, McGraw Hill
 \bibitem[]{}
   Broten, N.W., MacLeod, J.M., Avery, L.W. et al., 1984, ApJ 276, L25
 \bibitem[]{}
   Chin, Y.-N., 1995, Ph.D. Thesis, Bonn University
 \bibitem[]{}
   Chin, Y.-N., Henkel, C., Whiteoak, J.B., Langer, N., Churchwell, E.B.,
     1996, A\&A 305, 960
 \bibitem[]{}
   Cummins, S.E., Green, S., Thaddeus, P., Linke, R.A., 1983, ApJ 266, 331
 \bibitem[]{}
   Frerking, M.A., Wilson, R.W., Linke, R.A., Wannier, P.G., 1980, ApJ 240, 65
 \bibitem[]{}
   Helmich, F.P., Jansen, D.J., de Graauw, T., Groesbeck, T.D.,
     van Dishoeck, E.F., 1994, A\&A 283, 626
 \bibitem[]{}
   Henkel, C., Wilson, T.L., Langer, N., Chin, Y.-N., Mauersberger, R., 1994,
     in: The structure and content of molecular clouds -- 25 years of molecular
     radioastronomy, eds.: T.L. Wilson, K.J. Johnston, Springer Verlag, p.72
 \bibitem[]{}
   Langer N., Arcoragi J.-P., Arnould M., 1989, A\&A 210, 187
 \bibitem[]{}
   Linke, R.A., Goldsmith, P.F., 1980, ApJ 235, 437
 \bibitem[]{}
   Lovas, F.J., 1984, SLAIM Magnetic Tape version T84, priv. comm.
 \bibitem[]{}
   Lovas, F.J., 1992, J. Phys. Chem. Ref. Data 21, 181
 \bibitem[]{}
   Thielemann F.-K., Arnett W.D., 1985, ApJ 295, 604
 \bibitem[]{}
   Wilson, T.L., Rood, R.T., 1994, ARA\&A 32, 191
 \bibitem[]{}
   Woosley, S.E., Arnett, W.D., Clayton, D.D., 1973, ApJS 26, 231
 \bibitem[]{}
   Woosley S.E., Weaver T.A., 1995, ApJS 101, 181
\end{thebibliography}
\end{document}